\documentclass[letterpaper, 10 pt, conference]{ieeeconf}
\usepackage{amsfonts}
\usepackage{epsfig}
\usepackage[cmex10]{amsmath}
\usepackage{color}
\usepackage{epsfig}
\usepackage{url}
\usepackage{cite}
\usepackage{enumerate}
\usepackage{epstopdf}
\usepackage{amssymb}
\usepackage{graphicx}
\usepackage{amsmath}
\usepackage{graphicx}
\usepackage{marvosym}
\usepackage{steinmetz}
\usepackage{xcolor}
\usepackage{cancel}
\usepackage {tikz}
\newtheorem{lem}{Lemma}
\newtheorem{rem}{Remark}

\allowdisplaybreaks
\newcommand{\norm}[1]{\left\lVert#1\right\rVert}

\begin{document}

\title{Secure State Estimation for Nonlinear Power Systems under Cyber Attacks}

\author{Qie Hu*, Dariush Fooladivanda*, Young Hwan Chang*, and Claire J. Tomlin
\thanks{This work is supported by NSF under CPS:ActionWebs (CNS-0931843) and CPS:FORCES (CNS1239166), by ONR under MIT-5710002646 (SMARTS MURI).}
\thanks{The first and last authors are with the Department of Electrical Engineering and Computer Sciences, University
of California Berkeley, USA (\{qiehu, tomlin\}@berkeley.edu). D. Fooladivanda is with the Department of Electrical and Computer Engineering, University of Illinois Urbana-Champaign, USA (dfooladi@illinois.edu). Y. H. Chang is with the Department of Biomedical Engineering, Oregon Health and Science University, USA (chanyo@ohsu.edu).}
\thanks{*These authors contributed equally to this work.}}


\maketitle
\begin{abstract}
This paper focuses on securely estimating the state of a nonlinear dynamical system from a set of corrupted measurements. In particular, we consider two broad classes of nonlinear systems, and propose a technique which enables us to perform secure state estimation for such nonlinear systems. We then provide guarantees on the achievable state estimation error against arbitrary corruptions, and analytically characterize the number of errors that can be perfectly corrected by a decoder. To illustrate how the proposed nonlinear estimation approach can be applied to practical systems, we focus on secure estimation for the wide area control of an interconnected power system under cyber-physical attacks and communication failures, and propose a secure estimator for the power system. Finally, we numerically show that the proposed secure estimation algorithm enables us to reconstruct the attack signals accurately.
\end{abstract}


\section{Introduction}


Today's critical infrastructures are mostly managed by cyber-physical systems (CPS) that consist of several actuators, sensors, and controllers. Securing these systems against malicious attacks or communication failures is an important problem \cite{cps1}. In recent years, researchers have investigated various aspects of the problem of securing complex CPS, e.g., the networking security among cyber devices \cite{security_0}-\!\!\cite{comm_net_4} and the early detection of faults/attacks \cite{comm_early_1}-\!\!\cite{comm_early_2}. From the controller's point of view, researchers have studied how we can securely estimate the state of a dynamical system from a set of noisy and maliciously corrupted sensor measurements. In particular, researchers have focused on linear dynamical systems, and have tried to understand how the system dynamics can be leveraged for security guarantees.

In \cite{ctrl_sec6}, Pasqualetti \textit{et al.} propose a mathematical framework for cyber-physical systems, attacks, and monitors. They study the fundamental monitoring limitations for linear systems (especially for power networks), and characterize the vulnerability of linear systems to cyber-physical attacks using graph theory. Finally, to detect and to identify attacks, the authors propose centralized and decentralized filters, and validate their results via several examples. The main limitation of the study is that the proposed filters are computationally expensive, and are hard to implement in practice. In \cite{ctrl_sec8}, Fawzi \textit{et al.} focus on secure estimation and control of linear time-invariant systems, and assume that the set of attacked nodes does not change
over time. The authors then formulate the system under attack as an estimation problem without any limiting assumption on attack signals, and propose a novel method for error estimation and correction. The main drawback of the study is that the set of attacked nodes is assumed to be fixed. Chang \textit{et al.} \cite{ctrl_sec9} extend the results in \cite{ctrl_sec8} to scenarios in which the set of attacked nodes
can change over time, and show that under a certain condition, the secure estimation problem with time-varying attacked nodes
is equivalent to the classical error correction problem. The authors provide a novel method to guarantee accurate decoding, and then propose a secure estimation method which is a combination of the proposed secure estimator and the Kalman Filter (KF). Finally, they demonstrate the performance of their algorithm through numerous simulations. 


Extensive work has been done on secure estimation for linear dynamical systems, here, we focus on the design of a secure state estimator for nonlinear systems. We study the secure estimation of the state of a nonlinear dynamical system from a set of corrupted measurements. In particular, we focus on two broad classes of nonlinear systems, and propose a technique which enables us to transform the nonlinear dynamics into a set of linear equations. We then apply the classical error correction method to the equivalent linear system, and provide guarantees on the achievable state estimation error against arbitrary corruptions. In this paper, we do not assume the sensor attacks or corruptions to follow any particular model. The only assumption concerning the corrupted sensors is about the number of sensors that are corrupted due to attacks or failures. Our analytical results characterize the number of errors that can be perfectly corrected by an estimator.

As mentioned earlier, we focus on two broad classes of nonlinear systems in this paper. In order to illustrate how the proposed nonlinear state estimation approach can be applied to practical systems, we consider an interconnected power system comprising several synchronous generators, transmission lines, and energy storage units. We assume that all these physical devices are controlled via wide area control systems (WACS) as well as local controllers which use the synchrophasor technology to maintain the system's stability. WACS employ advanced data acquisition, communications, and control to enable increased efficiency and reliability of power delivery \cite{pmu_w_0}-\!\!\cite{pmu_w_3}. While WACS are the most promising technology to detect small signal instabilities in large power systems, their performance is highly dependent on the received data and on the underlying communication network. Clearly, the greater dependence on communication systems increases opportunities for cyber-physical attacks and disturbances. Extensive work has been done on monitoring and on autonomous feedback control for WACS \cite{wacs_ref8}, but this work has not studied how to identify cyber-physical attacks or communication failures, and how to perform secure state estimation for WACS.

In this paper, we focus on secure estimation for the wide area control of an interconnected power system, and assume that the system operator has installed several phasor measurement units (PMU) at different generator buses in the power network. These PMUs are connected through a communication network which sends PMU measurements, including rotors' speeds and generators' phase angles, to different controllers in the system. We also assume that there are several communication channels between generators and the WACS, as well as channels between generators. These channels and PMUs are not secured and are subject to cyber attacks and failures (except the communication channels from the WACS to the generators). Therefore, the WACS needs to perform secure state estimation to reconstruct the system's states before using the received data for computing wide area control signals, and to monitor the operation of local controllers installed in the system. We develop a secure state estimator for the WACS by using the proposed secure state estimation technique for nonlinear systems.

The paper is organized as follows: In Section \ref{sec_review}, we review the classical error correction methods. We then formulate the nonlinear state estimation problem and propose a solution technique for the proposed problem in Section \ref{sec_nonlinear_estimation}. We illustrate how the proposed secure state estimation approach can be applied to practical systems in Section \ref{sec_application} and \ref{sec_application1}. Finally, we provide a numerical example in Section \ref{sec_example}.


\section{Error Correction: A Review}\label{sec_review}
\subsection{Classical Error Correction}
\textbf{Compressed Sensing:} Sparse solutions $x\in \mathbb{R}^n$, are sought to the following problem:
\begin{equation}
	\min_x \norm{x}_0 \text{ subject to } b= Ax
	\label{eq:CS}
\end{equation}
where $b \in \mathbb{R}^m$ are the measurements, $A \in \mathbb{R}^{m\times n}~ (m \ll n)$ is a sensing matrix and $\norm{x}_0$ denotes the number of nonzero elements of $x$. The following lemma provides a sufficient condition for a unique solution to (\ref{eq:CS}):

\begin{lem} (\hspace{1sp}\cite{Candes_Tao}) \label{lem:CS}
If the sparsest solution to (\ref{eq:CS}) has $\norm{x}_0 = q$ and $m\ge 2q$ and all subsets of $2q$ columns of $A$ are full rank, then the solution is unique.
\end{lem}
\begin{pf}
Suppose the solution is not unique, hence there exists $x_1 \neq  x_2$ such that $Ax _1 = b$ and $Ax_2 = b$ where $\norm{x_1}_0 = \norm{x_2}_0 = q$. Then $A(x_1 - x_2) = 0$ and $x_1 - x_2 \neq 0$. Since $\norm{x_1-x_2}_0 \leq 2q$ and all $2q$ columns of $A$ are full rank (i.e. linearly independent), it is impossible to have $x_1-x_2\neq 0$ that satisfies $A(x_1-x_2) = 0$ (contradiction).
\end{pf}

\textbf{The Error Correction Problem \cite{Candes_Tao}:} 
Consider the classical error correction problem: $y=Cx + e$ where $C\in \mathbb{R}^{l\times n}$ is a coding matrix $(l > n)$ and assumed to be full rank. We wish to recover the input vector $x \in \mathbb{R}^n$ from corrupted measurements $y$. Here, $e$ is an arbitrary and unknown sparse error vector. To reconstruct $x$, note that it is obviously sufficient to reconstruct the vector $e$ since knowledge of $Cx + e$ together with $e$ gives $Cx$, and consequently $x$ since $C$ has full rank \cite{Candes_Tao}. In \cite{Candes_Tao}, the authors construct a matrix $F$ which annihilates $C$ on the left, i.e.,  $FCx = 0$ for all $x$. Then, they apply $F$ to the output $y$ and obtain
\begin{equation}
	\tilde y = F (Cx + e) = Fe.
\end{equation}
Thus, the decoding problem can be reduced to that of reconstructing a sparse vector $e$ from the observations $\tilde y = Fe$. Therefore, by Lemma \ref{lem:CS}, if all subsets of $2q$ columns of $F$ are full rank, then we can reconstruct any $e$ such that $\| e \|_0 \leq q$.

\subsection{Secure Estimation via Error Correction (Linear Systems) \cite{ctrl_sec9} }


Consider the linear control system as follows:
\begin{equation}
\begin{aligned}
x(k+1)= A x(k) , ~  y(k) = C x(k) + e(k)
\end{aligned}
 \label{eq:system_model_se}
\end{equation}
where $x(k) \in \mathbb{R}^n $ and  $y(k) \in \mathbb{R}^p$ represent the states and outputs of the system at time $k$, respectively. $e(k) \in \mathbb{R}^p$ represents attack signals injected by malicious agents at the sensors, and the set of attacked sensors can change over time.

Consider the problem of reconstructing the initial state $x(0)$ of the plant from the corrupted observations $y(k)$'s where $k=0,...,T-1$.
Let $E_{q,T}$ denote the set of error vectors $\begin{bmatrix} e(0); ~ ...~  ;  e(T-1) \end{bmatrix}   \in  \mathbb{R}^{p\cdot T} $ where each $e(k)$ satisfies $\|e(k)\|_0 \leq q \leq p$. 
\begin{eqnarray} \label{eq:sys_err_corr}
\begin{aligned}
	Y &\triangleq \begin{bmatrix} y(0) \\ y(1) \\ \vdots \\ y(T-1) \end{bmatrix}
		= \begin{bmatrix} Cx(0) + e(0)\\ CA x(0) + e(1) \\ \vdots \\ CA^{T-1} x(0) + e(T-1) \end{bmatrix} \\
		& =
		\begin{bmatrix} C \\ CA \\ \vdots \\ CA^{T-1} \end{bmatrix} x(0) + E_{q,T} \triangleq \Phi x(0) + E_{q,T}
		\label{eq:decoder_Phi}
\end{aligned}
\end{eqnarray}
where $Y \in \mathbb{R}^{p\cdot T}$ is a collection of corrupted measurements over $T$ time steps and $\Phi \in \mathbb{R}^{p\cdot T \times n}$ represents the observability matrix of the system. 
We assume that $\text{rank}(\Phi) = n$. This is a reasonable assumption because if not, the system is unobservable, therefore we cannot determine $x(0)$ even if there was no attack ($E_{q,T} = 0$).

Inspired by error correction techniques proposed in \cite{Candes_Tao} and \cite{David_Chang2}, we first determine the error vector $E_{q,T}$, and then solve for $x(0)$. 
Consider the $QR$ decomposition of $\Phi \in \mathbb{R}^{p\cdot T \times n}$,
\begin{eqnarray}
	\Phi = \begin{bmatrix} Q_1 & Q_2 \end{bmatrix} \begin{bmatrix} R_1 \\ 0 \end{bmatrix} = Q_1 R_1
\end{eqnarray}
where $\begin{bmatrix} Q_1 & Q_2 \end{bmatrix} \in \mathbb{R}^{p\cdot T \times p\cdot T}$ is orthogonal, $Q_1 \in \mathbb{R}^{p\cdot T\times n}, Q_2 \in \mathbb{R}^{p\cdot T \times (p\cdot T-n)}$, and $R_1 \in \mathbb{R}^{n\times n}$ is a rank-$n$ upper triangular matrix.
Pre-multiplying (\ref{eq:decoder_Phi}) by $\begin{bmatrix} Q_1 & Q_2 \end{bmatrix} ^\top$ gives:
\begin{equation}
	\begin{bmatrix} Q_1 ^\top \\ Q_2 ^\top \end{bmatrix} Y = \begin{bmatrix}R_1 \\ 0  \end{bmatrix} x(0) + \begin{bmatrix} Q_1 ^\top \\ Q_2^\top \end{bmatrix} E_{q,T}.
	\label{eq:QR}
\end{equation}
We can solve for $E_{q,T}$ using the second block row:
\begin{equation}
	\tilde Y \triangleq Q_2^\top Y = Q_2^\top E_{q,T}
	\label{eq:E_est}
\end{equation}
where $Q_2^\top \in \mathbb {R} ^{ (p\cdot T-n) \times p\cdot T}$.
From Lemma \ref{lem:CS}, (\ref{eq:E_est}) has a unique, $s (\le q\cdot T)$-sparse solution if all subsets of $2s(\le2 q\cdot T)$ columns of $Q_2^\top$ are full rank (this is a reasonable assumption if $ (p\cdot T-n) \ge 2s = 2q\cdot T$). Thus, we consider solving the following $l_1$-minimization problem:
\begin{equation}
	\hat{E}_{q,T} = \arg \min_E \norm { E}_{l_1} \text{ subject to } \tilde Y = Q_2^\top E
	\label{eq:solve_E}
\end{equation}

Given $\hat{E}_{q,T}$, we can then solve for $x(0)$ from the first block row of (\ref{eq:QR}):
\begin{equation}
	x(0) = R_1^{-1} Q_1^\top (Y- \hat{E}_{q,T})
	\label{eq:QR1}
\end{equation}
The conditions for the existence of a unique solution are stated in the following Lemma.
\begin{lem} (\hspace{1sp}\cite{ctrl_sec9}) \label{lem:EC}
	$x(0) $ is the unique solution if all subsets of $2s$ columns of $Q_2 ^\top$ are linearly independent and $\Phi$ is full column rank. Also, this condition is equivalent to $\| \Phi z \|_0 > 2s$ 
for all $z \in \mathbb{R}^n \backslash \{ 0 \}$.
\end{lem}
\begin{pf}
By Lemma \ref{lem:CS} and noting that by definition the null space of $Q_2^\top$ equals the column space of $\Phi$. For the second statement, we refer to the proof of Proposition 2 in \cite{ctrl_sec9}.
\end{pf}

\section{Secure Estimation for nonlinear systems}\label{sec_nonlinear_estimation}
\subsection{Problem Formulation}
Consider a nonlinear dynamical system given by
\begin{equation}\label{eq:sys_dynamics}
\begin{aligned}
	x(k+1) &= A x (k) + f\big(x(k),e(k)\big) + u(k) \\
	y(k) &= C x(k) + e(k)
\end{aligned}
\end{equation}
where $x(k)\in \mathbb{R}^n$ represents the state at time $k\in
\mathbb{N}$, $A \in \mathbb{R}^{n\times n}$, $f(x(k),e(k)): \mathbb{R}^n \times \mathbb{R}^p \rightarrow \mathbb{R}^n$ represent the system's dynamics, and $u(k) \in \mathbb{R}^n$ is a control input. Also, $C \in \mathbb{R}^{p\times n}$ is the sensors' measurement matrix, $y(k) \in \mathbb{R}^p$ are the corrupted measurements at time $k$, and $e(k)\in \mathbb{R}^p$ represents attack signals injected by malicious agents at the sensors. If sensor $i\in \{1,2,\cdots,p\}$ is not attacked, then necessarily the $i$-th element of $e(k)$ is zero; otherwise, the $i$-th element of $e(k)$ can take any value in $\mathbb{R}$ (i.e., we do not assume the errors $e(k)$ to follow any particular model). The only assumption concerning the corrupted sensors is the number of sensors that are attacked or corrupted due to failures. Our analytical results then characterize the number of errors that can be corrected by a decoder.

Under certain conditions which are explained at the end of the next subsection, the problems of reconstructing the state $x(k)$ or the initial condition $x(0)$ are equivalent. Therefore, in the following, we focus on the problem of reconstructing the initial state $x(0)$ for two classes of nonlinear systems.

\subsection{Existence of Mapping Function with Error Correction}\label{map_1}
Let us assume that there exists a mapping function $g\big(y(k)\big): \mathbb{R}^p \rightarrow \mathbb{R}^n$ such that
\begin{equation}\label{eq:mapping1}
g\big(y(k)\big)=f\big(x(k),e(k)\big)~.
\end{equation}
By using measurements $y(k)$ and this mapping function $g\big(y(k)\big)$, we can transform the nonlinear system in (\ref{eq:sys_dynamics}) into a linear system for which the error correction technique introduced in Section \ref{sec_review} can be used to reconstruct the initial state $x(0)$. From (\ref{eq:sys_dynamics}) and (\ref{eq:mapping1}), we have $g\big(y(k)\big)=x(k+1)-A x(k)-u(k)$ for all $k$, therefore, we can construct a vector $Y$ as follows:
\begin{equation}\label{map_1_error_correction}
\begin{aligned}
	Y&=\begin{bmatrix} y(0) \\ y(1) - C \big(g(y(0))+ u(0)\big)  \\
	y(2) - C \big( A g(y(0)) + A u(0) + g(y(1) ) + u(1) \big) \\ \vdots   \\
	y(T-1) - C \big(A^{T-2} g(y(0)) + A^{T-2} u(0) + \cdots \big) \end{bmatrix}
	\\&= \begin{bmatrix} C \\ CA \\ CA^2 \\  \vdots \\ CA^{T-1}  \end{bmatrix} x(0) + \begin{bmatrix} e(0) \\ e(1) \\ e(2) \\ \vdots  \\ e(T-1)\end{bmatrix} = \Phi x(0) + E
\end{aligned}
\end{equation}
where $E=[e(0);e(1);\cdots;e(T-1)]\in \mathbb{R}^{p \cdot T}$ is the set of error vectors, and $\Phi=[C;CA;CA^2;\cdots;CA^{T-1}]$.

We can now apply the error correction technique introduced in Section \ref{sec_review} to the linear system in (\ref{map_1_error_correction}). While the proposed technique is very useful in reconstructing the initial state $x(0)$, we might not always be able to find a mapping function $g(\cdot)$ such that (\ref{eq:mapping1}) is satisfied. Next, we use feedback linearization to generalize our result to a larger class of nonlinear systems.

\subsection{Feedback Linearization}
Let us assume that there exist mapping functions $g\big(y(k)\big)$ and $h_1\big(x(k)\big)$ (possibly nonlinear), and a linear map $h_2\big(e(k)\big)$ such that:
\begin{equation}\label{eq:form_feedback}
f\big(x(k),e(k)\big)= g\big(y(k)\big) + h_1\big(x(k)\big) + h_2\big(e(k)\big)	
\end{equation}
where $g\big(y(k)\big): \mathbb{R}^p \rightarrow \mathbb{R}^n$, $h_1\big(x(k)\big): \mathbb{R}^n \rightarrow \mathbb{R}^n$, and $h_2\big(e(k)\big): \mathbb{R}^p \rightarrow \mathbb{R}^n$ are non-zero.
The control input can be chosen such that $u(k)=-h_1 \big(x(k) \big)+v(k)$ which cancels out the nonlinear term $h_1\big(x(k)\big)$ and gives:
\begin{equation}
\begin{aligned}
	g\big(y(k)\big) &= x(k+1) - A x(k) - v(k) - h_2\big(e(k)\big).\nonumber
\end{aligned}
\end{equation}
We can now construct a vector $Y$ as follows:
\begin{equation}
\begin{aligned}
	Y=&\begin{bmatrix} y(0) \\ y(1) - C \big(g(y(0)) + v(0) \big) \\
	y(2) - C \big( A g(y(0)) + Av(0)  + g(y(1))  + v(1) \big)\\ \vdots \\ y(T-1) - C \big(A^{T-2} g(y(0)) + A^{T-2} v(0) + \cdots \big) \end{bmatrix}  \\
	=&~\Phi x(0) + E+
	\begin{bmatrix} 0 \\ C h_2 \big(e(0) \big) \\ C A h_2 \big(e(0) \big) + C h_2 \big(e(1) \big)  \\ \vdots  \\ C A^{T-2} h_2 \big(e(0) \big)  \cdots \end{bmatrix}.
\end{aligned}
\end{equation}
Since $h_2(\cdot)$ is a linear map (i.e., $h_2 \big(e(k) \big) = H e(k)$ where $H \in \mathbb{R}^{n \times p }$), thus:
\begin{equation}\label{linear_feed}
\begin{aligned}
Y &= \Phi x(0) + \Psi E
\end{aligned}
\end{equation}
where the matrices $\Phi \in \mathbb{R}^{p\cdot T \times n}$ and $\Psi \in \mathbb{R}^{p\cdot T \times p\cdot T}$ are as follows
\begin{equation}
\begin{aligned}
\Phi=\begin{bmatrix} C \\ CA \\ CA^2 \\ \vdots \\ CA^{T-1} \end{bmatrix},
\Psi = \begin{bmatrix} I & & & &   \\
CH & I & & \\
CAH & CH & I & \\
\ddots & \ddots  &  \ddots &  \ddots & \\
CA^{T-2} H & \cdots & \cdots & \cdots & I
\end{bmatrix}.\nonumber
\end{aligned}
\end{equation}
We can now apply the error correction method introduced in Section \ref{sec_review} to the linearized system in (\ref{linear_feed}) and reconstruct $x(0)$ if we satisfy the condition in Lemma \ref{lem:EC}.

Next, we consider an interconnected power system with several synchronous generators, and illustrate how the proposed nonlinear state estimation approach can be applied for the secure state estimation of practical systems.


\section{Application to Power System State Estimation}\label{sec_application}
Consider a power system comprising $N$ generators, and $q>N$ buses indexed by $g_1,\cdots,g_N$ and $b_{1},\cdots,b_q$, respectively. Let $b_1,\cdots,b_N$ be the generator terminal buses, each
one connected to exactly one generator, and let $b_{N+1},\cdots,b_q$ be the load buses. Let $\mathcal{B}$ and $\mathcal{V}$ denote the set of buses and transmission lines, respectively. We assume that the network topology is fixed and known (i.e., the neighbors of each bus are known), and that the corresponding graph $\mathcal{H}(\mathcal{B},\mathcal{V})$ is connected.

\textbf{Load buses:} Let $V_i$ and $\delta_i$ denote the magnitude and phase angle of the voltage phasor, respectively, at load bus $i$. Let $P_i^e$ be the total active power leaving bus $i$ (i.e., the real power drawn by the load
at bus $i$). $P_i^e$ can be computed by
\begin{align}
\label{load_bus}&P_i^e=\sum_{j\in\mathcal{B}}{{V_i V_j |y_{ij}|~\text{sin}(\delta_i-\delta_j+\phi_{ij})}}
\end{align}
where $y_{ij}=g_{ij}+\sqrt{-1} b_{ij}$ is the admittance of the line between buses $i$ and $j$, and $\phi_{ij}$ equals $\text{arctan}(g_{ij}/b_{ij})$. Note that $g_{ij}=g_{ji}\ge0$ and $b_{ij}=b_{ji}>0$ are the conductance and susceptance of the line between buses $i$ and $j$, respectively.

\textbf{Generator buses:} Let $\mathcal{G}$ denote the set of generators in the system. Let $\widehat{E_i}=E_i \phase{\theta_i}$ denote the internal voltage phasor of generator $i$. According to the synchronous machine theory, $E_i$ is constant and $\theta_i$ is the angular position of the generator rotor. We assume that the voltages at the generator buses are controlled via droop control, and that all the generator terminal buses are equipped with fast response energy storage units which are controlled via local and wide area controllers. Under these assumptions, at each generator $i\in\mathcal{G}$, the dynamic variables are the generator phase angle $\theta_i$ and the rotor speed $\omega_i$, and the generator dynamics can be described by \cite{Kundur}
\begin{align}
\label{swing1k}&\dot{\theta}_i=\omega_i-\omega_s\\
\label{swing2k}&\frac{2H_i}{\omega_s} \dot{\omega}_i=P_i^m-P_i^e-\frac{d_i}{\omega_s} \omega_i+ U_i
\end{align}
where $\omega_s$ is the nominal speed, $H_i$ is the machine inertia constant, $d_i$ is the damping coefficient of the generator, $U_i$ is the external stabilizing energy source at generator bus $b_i$, and $P_i^m$ is the mechanical power input to the generator. As mentioned earlier, each generator terminal bus $b_i$ is equipped with a fast response energy storage, such as flywheels, to improve the system stability. The energy storage receives a measurement-based control signal, and supplies or absorbs $U_i$ into bus $b_i$. Similar to the study in \cite{facts_storage2}, we develop a feedback linearization controller, and assume that generator $i$ implements the following feedback linearization control law
\begin{align}\label{Eq:Ui}
U_i = & - P^m_{i} + P_{i,\text{meas}}^e - F_i \left(\omega_i -\omega_s\right).
\end{align}
where $P_{i,\text{meas}}^e$ is computed locally at generator $i$ using received phase angle measurements and $F_i \geq 0$ is a design parameter. The energy storage injects $U_i$ per unit values of power into generator bus $b_i$ if $U_i\ge0$; otherwise, it absorbs $U_i$ per unit values of power from generator bus $b_i$.

\begin{figure}
\includegraphics[width=0.50\textwidth]{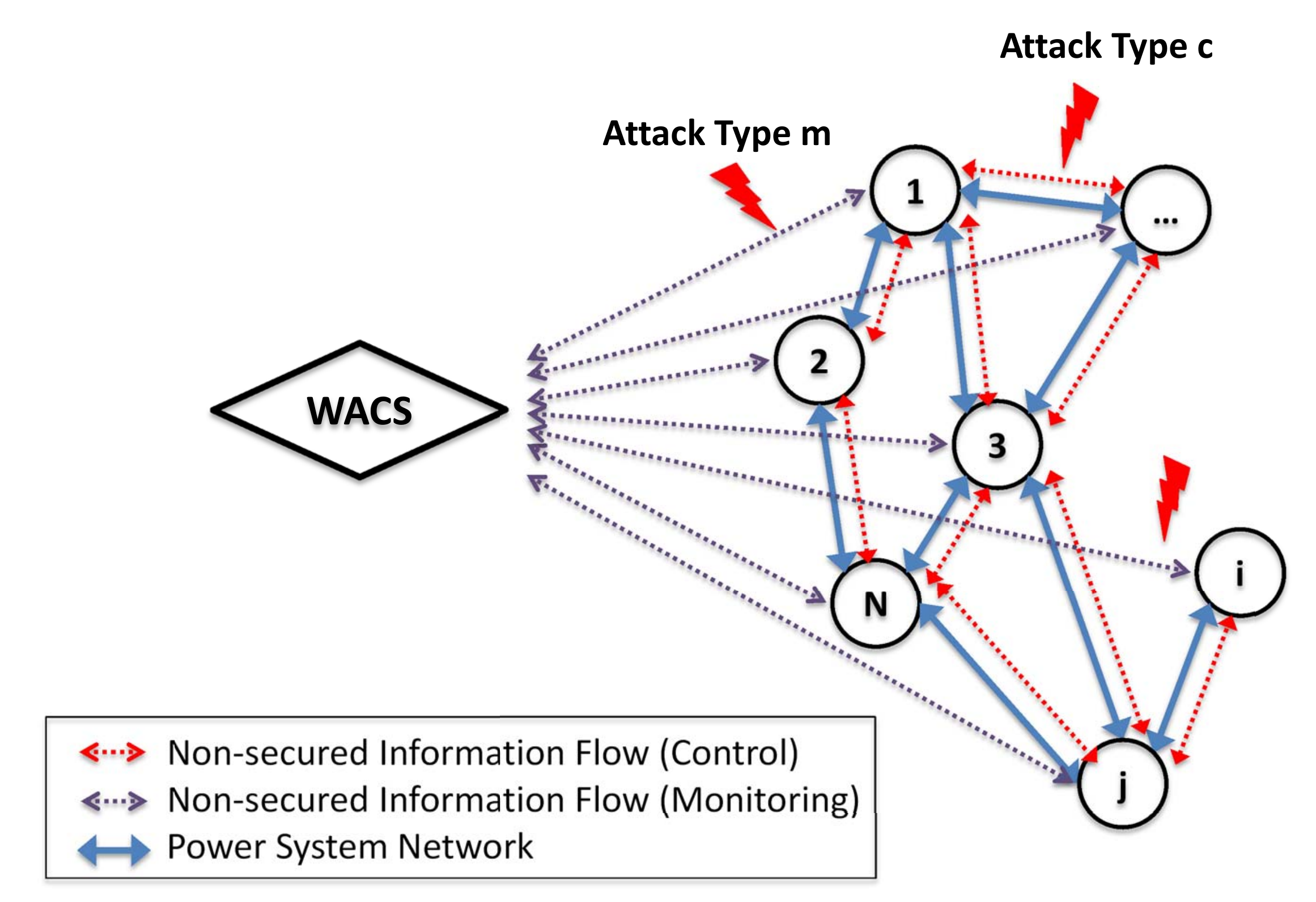} \caption{The networked system: the communication network and the power system.}
\label{wacs}
\end{figure}

\textbf{Wide Area Control System (WACS):} To maintain the system's stability, the system operator installs several PMUs at different buses in the power network. These PMUs are connected through a communication network which sends PMU measurements, including rotors' speeds and generators' phase angles, to different controllers in the system. We assume that there are several communication channels between generators and the WACS, as well as channels between generators as shown in Fig. \ref{wacs}. These channels and PMUs are not secured and are subject to cyber attacks and failures (except the communication channels from the WACS to the generators). Therefore, the WACS needs to perform secure state estimation before using the received data (e.g., $\omega_i$'s and $\theta_i$'s) for computing wide area control signals and for monitoring local controllers. Next, by using the proposed secure state estimation technique, we develop a secure state estimator for the WACS.

\section{Secure State Estimator for Wide Area \\Control Systems}\label{sec_application1}
We begin with deriving the system dynamics. The algebraic-differential equations (\ref{load_bus})-(\ref{swing2k}) describe system dynamics and power flows in the power system. However, we can use the Kron reduction technique to reduce the model in (\ref{load_bus})-(\ref{swing2k}) into a purely differential model, i.e., a network of electro-mechanical oscillators comprising the $N$ generators connected to each other through $L$ transmission lines. Let $\mathcal{V}^\prime$ denote the set of transmission lines between the generators after performing the Kron reduction technique, and let $\mathcal{K}(\mathcal{G},\mathcal{V}^\prime)$ denote the corresponding graph. This graph is connected and has $L$ edges where $L \le N (N-1)/2$.

Let $\widehat{y}_{ij}=\widehat{g}_{ij}+\sqrt{-1} ~\widehat{b}_{ij}$ denote the admittance of the the Kron-reduced equivalent line between generators $i$ and $j$, and let $\mathcal{N}_i$ denote the set of neighbors of generator $i$ after performing the Kron reduction technique. The system dynamics can be described by
\begin{equation}
\begin{aligned}
\dot{\theta_i} & =\omega_i-\omega_s\\
\frac{2H_i}{\omega_s} \dot{\omega_i} &=P_i^m-\sum_{j\in\mathcal{N}_i}{{E_i E_j {|\widehat{y}_{ij}|} ~ \text{sin}(\theta_i-\theta_j+\widehat{\phi}_{ij})}}\\
&\quad-\frac{d_i}{\omega_s} \omega_i+ U_i
\end{aligned}
\end{equation}
where $\widehat{\phi}_{ij}$ equals $\text{arctan}(\widehat{g}_{ij}/\widehat{b}_{ij})$. In this study, we consider the mechanical power input $P_i^m$ and the storage control signal $U_i$ as local control signals which are computed based on PMU measurements and wide area control signals (e.g., area control error) \cite{Kundur}. More precisely, the WACS performs a monitoring role and does secure estimation.


If measurements are attacked $P_{i,\text{meas}}^e \neq P_{i}^e$, then $\omega_i \neq \omega_{s}$. The WACS estimates attacks from measurements and communicates estimated attack signals to each generator. Then, each generator subtracts the estimated attack from the measured $\theta$ values, as to make $P_{i,\text{meas}}^e$ as close to $P_{i}^e$ as possible, and hence stabilize $\omega_i$ to $\omega_{s}$.

\subsection{Formulation of Secure Estimation}
In order to compute $P_{i,\text{meas}}^e$ and the local control signal $U_i$, each generator $i$ receives measurement $y^c_{ij}$ from its neighbor $j$ at time instant $t$:
\begin{equation}\label{Eq:yc}
\begin{aligned}
y^c_{ij}(t) & = \theta_j(t) + e^c_{ij} (t),~ i \in \mathcal{G},~j \in \mathcal{N}_i \cup \{i\}, \,
\end{aligned}
\end{equation}
where $e^c_{ij}$ represents the corruption in the measurement of $\theta_j$ received at generator $i$. Then, $P_{i,\text{meas}}^e$ can be computed by
\begin{equation}
P_{i,\text{meas}}^e = \sum_{j\in\mathcal{N}_i}{E_i E_j {|\widehat{y}_{ij}|} ~ \text{sin}\big(y^c_{ii}(t)-y^c_{ij}(t)+\widehat{\phi}_{ij}\big)}.
\end{equation}
Note that $e^c_{ji}$ represents the corruption in the measurement of $\theta_i$ received at generator $j$, thus $e^c_{ij}$ and $e^c_{ji}$ represent two distinct variables. In addition, $e^c_{ii} = 0$ because $\theta_i$ is measured locally and therefore is not subject to cyber attack, i.e., $y^c_{ii}(t) = \theta_i(t)$.

To enable the WACS to perform secure estimation, each generator sends all of its measurements to the WACS. In other words, the WACS receives the following measurements
\begin{equation}\label{Eq:ym}
\begin{aligned}
y^m_{ij}(t) &= y^c_{ij}(t) + e^m_{ij}(t),~i \in \mathcal{G},~j \in \mathcal{N}_i \cup \{i\} \,
\end{aligned}
\end{equation}
where $e^m_{ij}$ represents the corruption in $y^c_{ij}$.

We now apply the forward Euler discretization scheme to this continuous-time system and obtain the following discrete-time approximation, assuming a constant discretization step $T_s$ for all $k$:
\begin{equation}
\begin{aligned}
\theta_i (k+1) &= \theta_i(k) + T_s \big(w_i(k) - w_s\big) \\
w_i(k+1)
	&= \alpha w_i(k) + \beta \\
	& \quad + \sum_{j\in \mathcal{N}_i} f_{ij} \big(\theta_i(k), \theta_j(k), y^c_{ii}(k), y^c_{ij}(k)\big) \\
\end{aligned}
\end{equation}
where $\alpha = 1- \frac{T_s (d_i + F_i)}{2H_i}$, $\beta = \frac{T_s F_i w_s}{2H_i}$, $f_{ij} (\cdot) = \tilde G_{ij} \big[ \sin \big(\theta_i(k) - \theta_j(k) + \widehat{\phi}_{ij}\big) - \sin \big(y^c_{ii}(k) - y^c_{ij}(k)+\widehat{\phi}_{ij}\big) \big] $
and $\tilde G_{ij} = - \frac{T_s E_i E_j |\widehat{y}_{ij}|}{2H_i}$.

Using (\ref{Eq:yc}) and (\ref{Eq:ym}), $f_{ij} (\cdot)$ can be re-written using the measurements received at the WACS, $y^m_{ij}$'s:
\begin{equation} \label{Eq:case2_f_ij}
\begin{aligned}
	f_{ij} (\cdot ) &= \tilde G_{ij} \big[ \sin \big(\widehat{\phi}_{ij} + \theta_i(k) - \theta_j(k)\big) \\
		& \quad- \sin \big(\widehat{\phi}_{ij} + y^c_{ii}(k) - y^c_{ij}(k)\big) \big]\\
		& = \tilde G_{ij}  \sin \big(\widehat{\phi}_{ij} + y^m_{ii}(k) - y^m_{ij}(k) - e^m_{ii}(k) \\
		& \quad + e^m_{ij}(k) + e^c_{ij}(k)\big) - \tilde G_{ij} \sin \big(\widehat{\phi}_{ij} + y^m_{ii}(k) \\
		& \quad - y^m_{ij}(k) - e^m_{ii}(k) + e^m_{ij}(k)\big) \\
		& = G_{ij}^s (k) \epsilon_{ij}^c  (k)  -  G_{ij}^c  (k)  \epsilon_{ij}^s (k),
\end{aligned}\nonumber
\end{equation}
where $ G_{ij}^s(k) = \tilde G_{ij} \sin \big(\widehat{\phi}_{ij} + y^m_{ii}(k) - y^m_{ij}(k)\big)$, $ G_{ij}^c (k) = \tilde G_{ij} \cos \big(\widehat{\phi}_{ij} + y^m_{ii}(k) -y^m_{ij}(k)\big)$ are known to the WACS. On the other hand, $\epsilon_{ij}^c$ and $\epsilon_{ij}^s$ are functions of unknown attack signals and are defined as:
\begin{equation}\label{eq:epsilon_cs}
\begin{aligned}
\epsilon_{ij}^c(k) &= \cos \big(e^m_{ii}(k) - e^m_{ij}(k) - e^c_{ij}(k)\big) \\
& \quad - \cos \big(e^m_{ii}(k) - e^m_{ij}(k)\big) \\
\epsilon_{ij}^s (k) &= \sin \big(e^m_{ii}(k) - e^m_{ij}(k) - e^c_{ij}(k)\big) \\
& \quad - \sin \big(e^m_{ii}(k) - e^m_{ij}(k)\big).
\end{aligned}
\end{equation}
In other words, $f_{ij}(\cdot)$ is now a linear function of the unknowns: $\epsilon_{ij}^c(k)$ and $\epsilon_{ij}^s(k)$, whose coefficients can be computed by the WACS from its measurements. In addition, if there is no attack on any of the communication channels in the system at time slot $k$, then $\epsilon_{ij}^c (k) = \epsilon_{ij}^s (k) = 0$.

The state space model of the $i$-th generator is given by:
\begin{equation}
\begin{aligned}
x_i(k+1) &= A_i x_i(k) + q_i + H_i(k) \epsilon_i(k) \\
& = \begin{bmatrix} 1 & T_s \\ 0 & \alpha \end{bmatrix} x_i(k) + \begin{bmatrix} -T_s w_s \\ \beta \end{bmatrix} + \begin{bmatrix} 0 \\ h_i(k)^\top \end{bmatrix} \epsilon_i(k)
\end{aligned}
\end{equation}
where the state vector $ x_i(k) = \begin{bmatrix}  \theta_i(k), w_i(k) \end{bmatrix}^\top$ and
\begin{equation}
\begin{aligned}
h_i(k)^\top & = \big[ G_{i\mathcal{N}_i(1)}^s(k), \cdots, G_{i\mathcal{N}_i(l_i)}^s(k), \\
& \quad \quad \quad G_{i\mathcal{N}_i(1)}^c(k),  \cdots,  G_{i\mathcal{N}_i(l_i)}^c(k) \big] \in \mathbb{R}^{1 \times 2l_i}\\
\epsilon_i(k) & =  \big[\epsilon_{i\mathcal{N}_i(1)}^c(k), \cdots , \epsilon_{i\mathcal{N}_i(l_i)}^c(k), \\
& \quad\quad\quad \epsilon_{i\mathcal{N}_i(1)}^s (k), \cdots, \epsilon_{i\mathcal{N}_i(l_i)}^s (k) \big] ^\top \in \mathbb{R}^{2l_i \times 1}.
\end{aligned}\nonumber
\end{equation}
Here, $\mathcal{N}_i(j)$ is the $j$-th generator in the neighborhood of generator $i$ and $l_i$ is the cardinality of the set $\mathcal{N}_i$.

Now, consider the enlarged system with $N$ generators in the network:
\begin{equation}\label{eq:state_space_N}
\begin{aligned}
X(k+1) & = AX(k) + q + H(k) \epsilon(k) \\
Y(k) & = CX(k) + DE(k)
\end{aligned}
\end{equation}
where
\begin{equation}
\begin{aligned}
X(k) &= \begin{bmatrix} x_1(k); \cdots; x_N(k) \end{bmatrix} \in \mathbb{R}^{2N \times 1} \\
A &= \text{blkdiag} \{ A_1, \cdots, A_N \} \in \mathbb{R}^{2N \times 2N}\\
q &= \begin{bmatrix} q_1; \cdots; q_N  \end{bmatrix} \in \mathbb{R}^{2N \times 1}\\
H(k)& = \text{blkdiag} \{ H_1(k), \cdots, H_N(k) \}  \in \mathbb{R}^{2N \times 4L }\\
\epsilon(k) &\triangleq \begin{bmatrix} \epsilon_1(k); \cdots; \epsilon_N(k)\end{bmatrix} \in \mathbb{R}^{4L \times 1}  \\
Y(k) &= \begin{bmatrix} Y_{i}(k); \cdots; Y_N(k) \end{bmatrix} \in \mathbb{R}^{(N+2L) \times 1} \\
Y_i(k) & = \begin{bmatrix} y_{ii}(k); y_{i\mathcal{N}_i(1)}(k); \cdots; y_{i\mathcal{N}_i(l_i)}(k) \end{bmatrix} \in \mathbb{R}^{(1+l_i) \times 1} \\
D &= \begin{bmatrix} D_1, D_2 \end{bmatrix}  \in \mathbb{R}^{(N+2L)\times (N+4L)}\\
D_1 &= \text{blkdiag} \left \{ \begin{bmatrix} 0 \\ I_{l_1, l_1} \end{bmatrix}, \cdots, \begin{bmatrix} 0 \\ I_{l_N, l_N} \end{bmatrix} \right \} \in \mathbb{R}^{(N+2L) \times 2L} \\
D_2 & = I_{N+2L, N+2L} \in \mathbb{R}^{(N+2L) \times (N+2L)} \\
E(k) &=  \begin{bmatrix} E^c_1(k); \cdots; E^c_N(k); E^m_1(k) ; \cdots;  E^m_N(k) \end{bmatrix} \\
& \quad\in \mathbb{R}^{(N+4L) \times 1} \\
E^c_i(k) & = \begin{bmatrix} e^c_{i\mathcal{N}_i(1)}(k); \cdots; e^c_{i\mathcal{N}_i(l_i)}(k) \end{bmatrix} \in \mathbb{R}^{l_i \times 1} \\
E^m_i(k) & = \begin{bmatrix} e^m_{ii}; e^m_{i\mathcal{N}_i(1)}(k); \cdots; e^m_{i\mathcal{N}_i(l_i)}(k) \end{bmatrix} \in \mathbb{R}^{(1+l_i) \times 1}
\end{aligned}\nonumber
\end{equation}
and $L = \frac{\sum_i l_i}{2}$ represents the total number of edges / links in the network. Matrix $C \in \mathbb{R}^{(N+2L)\times 2N}$ is given as follows: let the $a$-th element of vector $Y$ be $y^m_{ij}$, then the $(a,b)$-th entry of $C$ is given by
\begin{equation}
C_{(a,b)} = \begin{cases}1 & \mbox{if } 2j-1 = b \\ 0 & \mbox{otherwise.} \end{cases} \nonumber
\end{equation}

\noindent
Consider $T$ time steps of measurements (i.e., $k = \{0, \cdots, T-1 \}$) and define:
\begin{equation}
\bar Y = \begin{bmatrix} Y (0)  \\ Y (1) - C q  \\ \vdots \\ Y(T-1) - C \sum_{m=0}^{T-2}A^{T-2-m} q \end{bmatrix} \in \mathbb{R}^{(N+2L)T \times 1}
\end{equation}
then
\begin{equation}\label{eq:Ebar_Psi}
\bar Y = \Phi X(0) + \Psi  \bar E
\end{equation}
where $\Phi = \begin{bmatrix} C; CA ; \cdots ; CA^{T-1} \end{bmatrix} \in \mathbb{R}^{(N+2L)T \times 2N}$ is the observability matrix of the system, $\bar E = \begin{bmatrix} E(0); \cdots; E(T-1); \epsilon(0); \cdots; \epsilon(T-2) \end{bmatrix} \in \mathbb{R}^{((N+4L)T + 4L(T-1)) \times 1} $ and $\Psi = \begin{bmatrix} \Psi_{1} & \Psi_{2} \end{bmatrix}$, with $\Psi_{1} \in \mathbb{R}^{(N+2L)T \times (N+4L)T}$ and $\Psi_{2}\in \mathbb{R}^{(N+2L)T \times 4L(T-1)}$ as follows:
\begin{equation}
\begin{aligned}
\Psi_{1} &= \text{blkdiag} \{ D, \cdots, D \} \\
\Psi_{2} &= \begin{bmatrix}   0 &  0  &  \cdots    	\\
   			        C H(0) & 0 &  \cdots  	\\
			       CAH(0) &  CH(1) & \cdots	\\
			        \vdots & \vdots  &  \ddots \\
			       CA^{T-2}H(0) & \cdots & CH(T-2) \\			
			    \end{bmatrix}	.
\end{aligned}\nonumber
\end{equation}

\noindent
We can choose $\Omega \in \mathbb{R}^{ ((N+2L)T - 2N )\times (N+2L)T}$ such that $\Omega \Phi = 0$, then:
\begin{equation}\label{eq:Ebar_OmegaPsi}
	\tilde Y = \Omega \bar Y = \Omega \Psi \bar E,
\end{equation}
where $\Omega \Psi \in \mathbb{R}^{((N+2L)T - 2N) \times ((N+4L)T + 4L(T-1))}$.


\subsection{Difficulties in Secure Estimation due to System Dynamics}
The linear system in (\ref{eq:Ebar_OmegaPsi}) is in the form of (\ref{eq:E_est}). Hence, from Lemma \ref{lem:CS}, $\bar E$ has a unique $s$-sparse solution if all subsets of $2s$ columns of  $\Omega\Psi$ are linearly independent. However, in this case, $\Psi_2$ is rank deficient due to the system dynamics. To see this more clearly, the first row of $H_i(k)$ are zeros, therefore given any matrix $M$, $\text{rank}(MH_i(k)) = 1$. In other words, all columns of $\Psi_2$ that correspond to the same $H_i(k)$ are linearly dependent. In (\ref{eq:Ebar_OmegaPsi}), $\Psi_2$ multiplies the $\epsilon(k)$ terms in $\bar E$, consequently, the $\epsilon(k)$'s may not be identifiable.

Furthermore, $y^m_{ij}(k) = \theta_j(k) + e^c_{ij}(k) + e^m_{ij}(k)$, hence, for a given ($i,j$)-pair ($i \neq j$) and a given $k$, the columns of $\Psi_1$ that correspond to $e^c_{ij}(k)$ and $e^m_{ij}(k)$ are identical. As a result, applying the estimation algorithm introduced in Section \ref{sec_review} to (\ref{eq:Ebar_OmegaPsi}) can only uniquely identify the sum $e^c_{ij}(k)+e^m_{ij}(k)$ but not the individual terms: $e^c_{ij}(k)$ and $e^m_{ij}(k)$.

To overcome these difficulties, we consider the following problem setup.


\subsection{Assumptions and Secure Estimation with 2-Step Delay}

We distinguish between two types of attacks:
\begin{itemize}
\item \textbf{c-attack:} an attack that corrupts communication channels between generators, i.e., there exists at least one ($i,j$)-pair such that $e^c_{ij} \neq 0$.
\item \textbf{m-attack:} an attack that affects communication channels between the generators and the WACS, i.e., there exists at least one ($i,j$)-pair such that $e^m_{ij} \neq 0$.
\end{itemize}
\noindent
Assume that at any time $k$, the power network is subject to either a c-attack or an m-attack, but not both.
However, both the type of attack and the set of attacked measurements can change at each time step.
In addition, the WACS does not know \textit{a priori} which type of attack the system is subjected to, hence secure estimation techniques are used to determine the type of attack, as well as the exact corruption signals.

From (\ref{eq:state_space_N}) we can derive the following equality:
\begin{equation}\label{eq:2step_delay}
\begin{aligned}
f_\text{2-step}\big (\epsilon(k-2), E(k) \big) & = Y(k) - CA^2\cdot X(k-2) - CAq  \\
	& \quad  - Cq - CA\cdot H(k-2)\cdot \epsilon(k-2)  \\
	& \quad - DE(k)\\
	& = 0,
\end{aligned}
\end{equation}
where the first equality uses $CH(k)=0$ for all $k$. Observe that if it is an m-attack at time $k$, then $e^c_{ij}(k) + e^m_{ij}(k) = e^m_{ij}(k)$ and $\epsilon(k) = 0$ in (\ref{eq:2step_delay}); on the other hand, if it is a c-attack at time $k$, then $e^c_{ij}(k) + e^m_{ij}(k) = e^c_{ij}(k)$ and $\epsilon(k) \neq 0$.
Combining these observations with (\ref{eq:2step_delay}), we propose the following algorithm which can be used by the WACS to determine the type of attack and the exact corruption signals, with a 2-step delay.

\noindent \textbf{Step~1:} At each time $k$, estimate $\bar E(k)$ by solving the following $l_1$-minimization problem
\begin{equation}
\bar E_\text{est}(k) = \arg \min \| \bar E \|_{l_1} ~\text{subject to}~ \tilde Y = \Omega\Psi \bar E.
\nonumber
\end{equation}
Extract $E_\text{est}(k-2)$ and $E_\text{est}(k)$ from $\bar E_\text{est}(k)$.

%
\noindent \textbf{Step~2:} Let $e^c_{\text{est},ij}$ and $e^m_{\text{est},ij}$ denote $e^c_{ij}$ and $e^m_{ij}$ elements in $E_\text{est}$. Similarly, for $E_{\text{c-att,est}}$ and $E_{\text{m-att,est}}$. 
Consider 2 cases:
\begin{itemize}
\item \textit{m-attack at time $k-2$}: $\epsilon_\text{m-att,est}(k-2) = 0$.
\item \textit{c-attack at time $k-2$}: 
First, obtain $E_\text{c-att,est}(k-2)$ as follows: for all $i$ and $j$, $e^m_{\text{c-att,est}, ij}(k-2) = 0$ and $e^c_{\text{c-att,est},ij}(k-2) = e^c_{\text{est},ij}(k-2) + e^m_{\text{est},ij}(k-2)$. Second, use (\ref{eq:epsilon_cs}) to compute $\epsilon_\text{c-att,est}(k-2)$ from $E_\text{c-att,est}(k-2)$.
\end{itemize}

\noindent \textbf{Step~3:} Evaluate equation (\ref{eq:2step_delay}) for both cases. If $\|f_\text{2-step}\big (\epsilon_\text{c-att,est}(k-2), E_\text{est}(k) \big) \| < \|f_\text{2-step}\big (\epsilon_\text{m-att,est}(k-2), E_\text{est}(k) \big) \|$, then it is a c-attack at $k-2$, and $E(k-2) = E_\text{c-att,est}(k-2)$, $\epsilon(k-2) = \epsilon_\text{c-att,est}(k-2)$. Otherwise, it is an m-attack at $k-2$, and $E(k-2) = E_\text{m-att,est}(k-2)$, $\epsilon(k-2) = \epsilon_\text{m-att,est}(k-2)$.

To summarize, as a result of the system dynamics and the proposed model, it is not possible to recover the exact corruption if the system is subjected to both c- and m-attacks at the same time. In light of this, we make the simplifying assumption that at any time $k$, the system may only be subject to one type of attack. However, the type of attack can change over time. Then, by comparing the actual measurements with the different system trajectories that would result from each type of attack, we can determine both the attack type and the exact corruption signals, with a 2-step delay. Note that at time $k$, this secure state estimation algorithm is able to detect the presence of attacks at times $k-1$ and $k$, merely not the exact attack signals.

\begin{rem}
Let $Q$ be the number of non-zero elements in $\bar E$ that correspond to the $E(k)$ terms.
By Lemma \ref{lem:EC}, the maximum number $Q$ that can be corrected after $T$ steps is $\lfloor (N+2L)T/2 - N \rfloor$, and the average number of correctable non-zeros per time step is $\bar q (= \frac {Q} {T}) = \lfloor N/2+L - N/T \rfloor$.
\end{rem}

Due to the non-identifiability of $e^c_ij(k)$ and $e^m_ij(k)$, each corrupted measurement at time $k$ contributes up to 2 non-zero elements in $E(k)$. Thus, in this example, the average number of corrupted measurements that can be perfectly recovered is $\lfloor N/4+L/2 - N/(2T) \rfloor$ per time step; and the maximum average number of correctable corruptions is $\lceil N/4+L/2 - 1\rceil$. Next, we numerically demonstrate the effectiveness of the proposed state estimation algorithm.




\section{Numerical Example}\label{sec_example}
\begin{figure}
\centering
\includegraphics[scale=0.46]{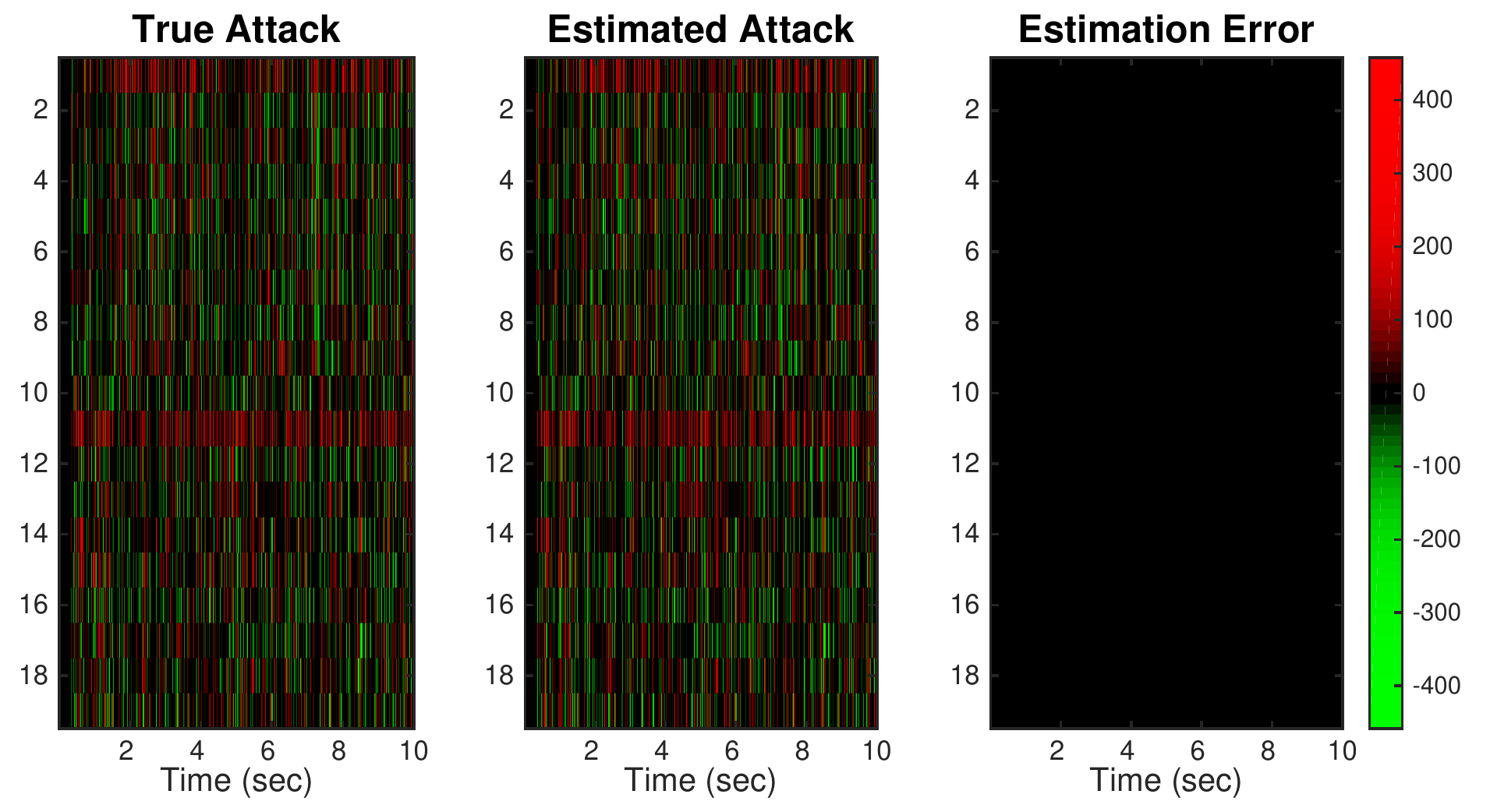}
\caption{True and estimated attack signals. The rows and columns correspond to attacked measurements and time steps, respectively. In subfigures ``True Attack'' and ``Estimated Attack'', the color indicates the attack signal: red is a positive attack, green a negative attack and black is no attack. In subfigure ``Estimation Error'', the black color indicates there is zero estimation error for all measurements at all times.}
\label{fig:attack}
\end{figure}

\begin{figure}
\centering
\includegraphics[scale=0.46]{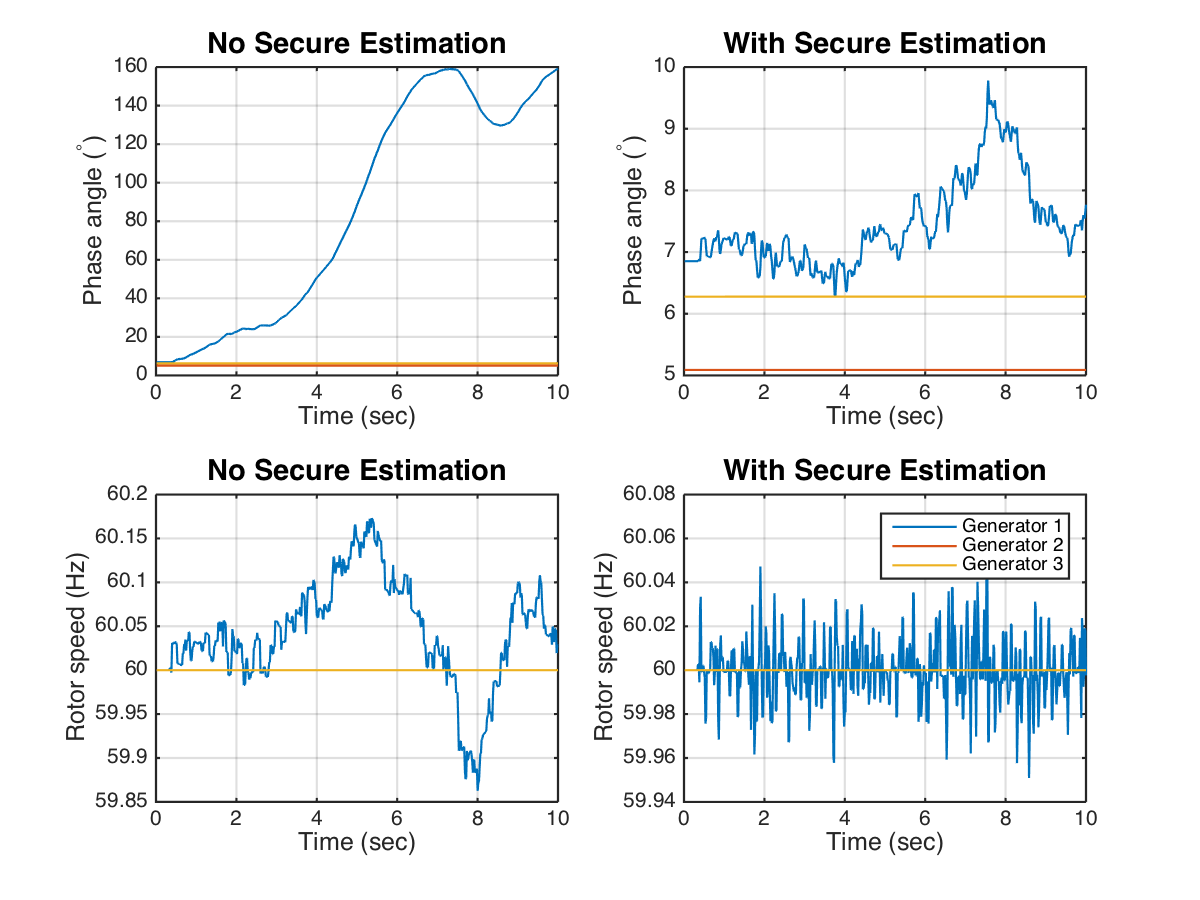}
\caption{Evolution of phase angle and rotor speed of generators 1, 2, and 3 when the network is under attack, with and without secure estimation. Without secure estimation, generator 1's phase angle deviates rapidly from its equilibrium value, reaching a maximum of $159^\circ$. With secure estimation, the phase angle of generator 1 is maintained close to equilibrium, with a maximum deviation of only $2.9^\circ$.}
\label{fig:SE}
\end{figure}

We focus on the New England power system comprising 10 generators and 39 buses. The values of the system parameters are taken from \cite{new_england1}-\!\cite{new_england2}. The power system is initially under equilibrium condition, and the rotor speeds of all 10 generators are at the nominal value, $w_s$, of 60~Hz. Malicious attacks targeted at generator 1 are injected as follows:
\begin{itemize}
\item At every time step, the attacker randomly chooses to perform either a c-attack or an m-attack. Denote the chosen attack type as $[\cdot]$.
\item A constant attack signal of $90^\circ$ is injected into the measurement of $\theta_2$, i.e., $e^{[\cdot]}_{12} = 90^\circ$.
\item 9 additional measurements from $E^{[\cdot]}_1$ are chosen at random, and each of them is corrupted with a random Gaussian signal.
\end{itemize}
The left plot in Figure \ref{fig:attack} shows the true attack signals (for clarity, only those measurements that are attacked at some point during the simulation are shown). Initially, generator 1 has a phase angle of $6.85^\circ$. The left plots in Figure \ref{fig:SE} show that if no secure estimation-based protection is implemented, the phase angle of generator 1 deviates rapidly from its equilibrium value and reaches $159^\circ$ within 7.3 sec. When the phase angle difference between two generators exceeds $90^\circ$, the generators can potentially loose synchrony and trip. As shown in Fig. \ref{fig:attack}, the proposed state estimation enables the WACS to reconstruct the rotor angles and rotational speeds of the generators accurately, and to prevent all possible failures.

We now consider the system's behavior with secure estimation-based protection. Using the algorithm described in Section \ref{sec_application1}, WACS estimates the attack signals and the generator phase angles. Then, it sends the reconstructed measurements to all the generators, and each generator uses the reconstructed phase angle measurements to compute its local control input.
During the simulation, the proposed estimation algorithm correctly estimates the type of attack at every time step (results are not shown due to space limitations) and also recovers the attack signals accurately as shown in Figure \ref{fig:attack}. Figure \ref{fig:SE} shows that by incorporating secure estimation in this way, even with a 2-step estimation delay, the phase angle of generator 1 is maintained within the range $[6.3^\circ,9.8^\circ]$, i.e., a maximum deviation of $2.9^\circ$ from its equilibrium value. In addition, generator 1's rotor speed hardly deviates from the nominal frequency of 60 Hz and is kept within $[59.95 \text{~Hz}, 60.07 \text{~Hz}]$. The phase angles of generators 4 to 10 are unaffected by the attack during the simulation period, and are not shown here for clarity.



\section{Conclusion}
We propose a secure state estimator for two broad classes of nonlinear systems, and provide guarantees on the achievable state estimation error against arbitrary corruptions. More precisely, we analytically characterize the number of errors that can be perfectly corrected by a decoder. We then illustrate how the proposed estimator can be used in practical systems. In particular, by using the proposed estimator, we propose a secure state estimator for the wide area control of power systems under cyber-physical attacks and communication failures. Finally, we numerically demonstrate the effectiveness of the proposed secure estimator.

\end{document}